\pdfoutput=1
\documentclass[sigconf,natbib,language=english,screen]{acmart}
\newif\ifacm
\acmtrue
\newif\ifarxiv
\arxivtrue

\usepackage[utf8]{inputenc}
\usepackage[english]{babel}

\ifacm
\usepackage{balance} %
\else
\usepackage{amsmath}
\usepackage{amsthm}
\usepackage{amsfonts}
\usepackage{amssymb}
\usepackage{amsopn}
\fi

\ifacm
\else
\usepackage{libertine}
\fi

\ifacm
\else
\usepackage{booktabs}
\fi
\usepackage{tabularx}

\usepackage{color}
\usepackage{bm}
\usepackage{lineno}
\usepackage{url}
\usepackage{hyperref}
\usepackage{cleveref}
\usepackage[boxruled,vlined,algo2e]{algorithm2e} 
\usepackage{setspace}
\usepackage{xcolor}
\usepackage{wrapfig}
\usepackage{graphicx}
\usepackage{pstool}
\usepackage{multirow}
\usepackage{setspace}
\usepackage{enumitem}
\usepackage{multirow}
\usepackage{subcaption}
\usepackage{comment}
\usepackage{xspace} %
\usepackage{tablefootnote} %

\graphicspath{{./figures/}}

\DeclareCaptionFormat{custom}
{%
    \textbf{#1#2}\textnormal{\small #3}
}
\usepackage{pgfplots}
\usepgfplotslibrary{external} 
\pgfplotsset{compat=newest}

\usepackage{tikz}%
\usetikzlibrary{shapes,snakes}
\usetikzlibrary{patterns}
\usetikzlibrary{shapes.misc}
\usetikzlibrary{arrows}
\usetikzlibrary{decorations.pathreplacing,calligraphy}
\usetikzlibrary{backgrounds}
\usetikzlibrary{external}
\usepgfplotslibrary{groupplots}
\usepgfplotslibrary{dateplot}

\newcommand{\code}[1]{\texttt{#1}}

\newcolumntype{Y}{>{\centering\arraybackslash}X}
\newcolumntype{Z}[1]{>{\hsize=#1\arraybackslash}X}

\newcommand{\be}{\begin{equation}}
\newcommand{\eec}{\;\;,\end{equation}}
\newcommand{\ee}{ \end{equation}}
\newcommand{\eed}{\;\;.\end{equation}}
\newcommand{\bse}{\begin{subequations}}
\newcommand{\ese}{\end{subequations}}

\newcommand{\Tref}[1]{\Cref{#1}}
\newcommand{\Fref}[1]{\Cref{#1}}

\newcommand{\Sref}[1]{\Cref{#1}}

\definecolor{darkGreen}{RGB}{0,160,0}
\definecolor{darkBlue}{RGB}{51,51,204}
\definecolor{darkRed}{RGB}{160,0,0}
\definecolor{orange}{RGB}{255,120,0}
\definecolor{myGreen}{rgb}{0.0706, 0.5333, 0.1921}

\newcommand{\revone}[1]{}
\newcommand{\revthree}[1]{}
\newcommand{\revtwo}[1]{}
\newcommand{\revall}[1]{}
\newcommand{\typo}[1]{}

\definecolor{airforceblue}{rgb}{0.36, 0.54, 0.66}
\definecolor{carolinablue}{rgb}{0.6, 0.73, 0.89}
\definecolor{moonstoneblue}{rgb}{0.45, 0.66, 0.76}
\definecolor{amber}{rgb}{1.0, 0.75, 0.0}
\definecolor{cadmiumorange}{rgb}{0.93, 0.53, 0.18}
\definecolor{persianorange}{rgb}{0.85, 0.56, 0.35}
\definecolor{cadetgrey}{rgb}{0.57, 0.64, 0.69}
\definecolor{darkgray}{rgb}{0.66, 0.66, 0.66}
\definecolor{asparagus}{rgb}{0.53, 0.66, 0.42}
\definecolor{cambridgeblue}{rgb}{0.64, 0.76, 0.68}
\definecolor{olivine}{rgb}{0.6, 0.73, 0.45}
\definecolor{carnelian}{rgb}{0.7, 0.11, 0.11}
\definecolor{cardinal}{rgb}{0.77, 0.12, 0.23}
\definecolor{firebrick}{rgb}{0.7, 0.13, 0.13}

\definecolor{matlabBlue}{rgb}{0, 0.4470, 0.7410}
\definecolor{matlabOrange}{rgb}{0.8500, 0.3250, 0.0980}
\definecolor{matlabYellow}{rgb}{0.9290, 0.6940, 0.1250}
\definecolor{matlabPurple}{rgb}{0.4940, 0.1840, 0.5560}
\definecolor{matlabGreen}{rgb}{0.4660, 0.6740, 0.1880}
\definecolor{matlabRed}{rgb} {0.7765, 0.1569, 0.0902}
\definecolor{matlabBlack}{rgb}{0, 0, 0}

\definecolor{keyBlue}{rgb}{0.0392, 0.2274, 0.4235}
\definecolor{keyBlueTrans}{rgb}{0.4274, 0.4980, 0.5764}
\definecolor{keyGreen}{rgb}{0.05490, 0.38039, 0.003921}
\definecolor{keyGreenTrans}{rgb}{0.7019, 0.8313, 0.7176}
\definecolor{keyOrange}{rgb}{0.98039, 0.4980, 0.03529}
\definecolor{keyGray}{rgb}{0.50196, 0.50196, 0.50196}
\definecolor{keyGreen2}{rgb}{0.0627 , 0.4196 , 0.3882}%
\definecolor{keyOrange2}{rgb}{0.9843 , 0.2941 , 0.2431}%

\definecolor{pyBlue}{rgb}{0.11372,0.42352,0.67059}
\definecolor{pyBlueLight}{rgb}{0.6196,0.79215,0.88235}
\definecolor{pyOrange}{rgb}{1.0,0.4549,0.0627}
\definecolor{pyOrangeLight}{rgb}{0.9921568,0.643137,0.376470}
\definecolor{pyGreen}{rgb}{0.1529,0.5882,0.1529}
\definecolor{pyRed}{rgb}{0.81569,0.13725,0.14117}
\definecolor{pyPurple}{rgb}{0.53725,0.36078,0.7098}
\definecolor{myCacaDoie}{rgb}{0.580392156862745,0.43,0.27}
\definecolor{pyBrown}{rgb}{0.549019607843137,0.337254901960784,0.294117647058824}
\definecolor{pyRose}{rgb}{0.890196078431372,0.466666666666667,0.76078431372549}
\definecolor{pyOrange2}{rgb}{0.9921,0.5098,0.2078}
\definecolor{pyBlue2}{rgb}{0.3765,0.6431,0.8157}
\definecolor{pyBar1}{rgb}{0.1647,0.2745,0.2745}
\definecolor{pyBar2}{rgb}{0.0000,0.4588,0.4588}
\definecolor{pyBar3}{rgb}{0.4863,0.7804,0.9998}
\definecolor{pyBar4}{rgb}{0.4549,0.4549,0.4549}

\tikzset{cross/.style={cross out, draw, minimum size=2*(#1-\pgflinewidth), inner sep=0pt, outer sep=0pt}}

\newcommand{\ColLegend}[2]{
        \tikz{
            \filldraw[draw=#1,fill=#2, line width=0.5 pt] (0pt,-3pt) rectangle (4pt,1pt);
        }
}

\definecolor{cadmiumorange}{rgb}{0.93, 0.53, 0.18}
\definecolor{asparagus}{rgb}{0.53, 0.66, 0.42}
\definecolor{cerulean}{rgb}{0.0, 0.48, 0.65}
\definecolor{brickred}{rgb}{0.8, 0.25, 0.33}
\definecolor{ferngreen}{rgb}{0.31, 0.47, 0.26}
\hypersetup{
    colorlinks=true,
    linkcolor=cerulean,
    filecolor=magenta,      
    urlcolor=brickred,
    citecolor=ferngreen
    }

\newcommand{\rev}[1]{#1}

\newcommand{\picfont}{\sffamily}
\pgfplotsset{
	reverse legend,
	legend style={anchor=north west,at={(0.01,0.99)},font=\tiny},
	legend image code/.code={
		\draw[very thick,
			 mark repeat=2,
			 mark phase=2,
			 dash phase=0pt,
			 ] plot coordinates {(0pt,0pt) (10pt,0pt)};%
	}
	}
\tikzset{font=\picfont{}}

\newcommand{\mygcd}[2]{\operatorname{gcd}\left( #1 \;,\; #2 \right)}

\newcommand{\npart}{N_{\text{part}}}

\newcommand{\spart}{S_{\text{part}}}

\newcommand{\ppt}{\theta}

\newcommand{\ptsingle}[1]{\code{Pt2Pt~single}#1}
\newcommand{\ptmany}[1]{\code{Pt2Pt~many}#1}
\newcommand{\ptpart}[1]{\code{Pt2Pt~part}#1}

\newcommand{\RMAsingle}[1]{\code{RMA~single~-~passive}#1}
\newcommand{\RMAmany}[1]{\code{RMA~many~-~passive}#1}
\newcommand{\RMAactive}[1]{\code{RMA~many~-~active}#1}
\newcommand{\RMAsingleactive}[1]{\code{RMA~single~-~active}#1}
\newcommand{\mpich}[1]{\code{MPICH}#1}

\newcommand{\openmp}[1]{\code{openmp}#1}
\newcommand{\vci}{{VCI}}

\newcommand{\ptp}{point-to-point}

\newcommand{\ssec}{\emph{s}}
\newcommand{\per}[1]{\,\!/\,\!}
\newcommand{\uspace}[1]{\,\!}
\newcommand{\B}{\emph{B}}
\newcommand{\GBs}{\emph{GB}\per{}\ssec}
\newcommand{\mus}{$\mu$\uspace{}\ssec}
\newcommand{\kB}{\emph{kB}}

\newcommand{\MB}{M\uspace{}B}

\newcommand{\eb}[1]{\emph{early-bird}#1}

\newcommand{\flop}[1]{\emph{flop}#1}
\newcommand{\ai}{\emph{AI}~}
\newcommand{\ci}{\emph{CI}~}

\definecolor{col_rma_single}{RGB}{214,39,40}
\definecolor{col_single}{RGB}{255,127,14}
\definecolor{col_rma_multi}{RGB}{227,119,194}
\definecolor{col_part}{RGB}{31,119,180}
\definecolor{col_multi}{RGB}{99,121,57}
\definecolor{col_rma_active}{RGB}{189,158,57}
\definecolor{col_rma_single_active}{RGB}{176,176,176}
\definecolor{col_stream}{RGB}{188,189,34}

\definecolor{col_model_black}{RGB}{176,176,176}
\definecolor{col_model_manyvci}{RGB}{255,127,14}
\definecolor{col_model_manyvcitwo}{RGB}{99,121,57}
\definecolor{col_model_onevci}{RGB}{31,119,180}

\newcommand{\CaptionSingle}{{\protect\ColLegend{none}{col_single}}}

\newcommand{\CaptionMulti}{{\protect\ColLegend{none}{col_multi}}}
\newcommand{\CaptionRMASingle}{{\protect\ColLegend{none}{col_rma_single}}}
\newcommand{\CaptionRMAMulti}{{\protect\ColLegend{none}{col_rma_multi}}}
\newcommand{\CaptionRMAActive}{{\protect\ColLegend{none}{col_rma_active}}}
\newcommand{\CaptionPart}{{\protect\ColLegend{none}{col_part}}}
\newcommand{\CaptionRMASingleActive}{{\protect\ColLegend{none}{col_rma_single_active}}}

\newcommand{\anl}{\affiliation{%
	\institution{Argonne National Laboratory}
	\streetaddress{9700 S Cass Ave}
	\city{Lemont}
	\state{Illinois}
	\country{USA}
	\postcode{60439}
}}

\title{Quantifying the Performance Benefits of\\ Partitioned Communication in {MPI}}
\author{Thomas Gillis}
\orcid{0000-0001-9740-4746}
\anl{}
\author{Ken Raffenetti}
\orcid{0009-0003-4705-2713}
\anl{}
\author{Hui Zhou}
\orcid{0000-0002-4422-2911}
\anl{}
\author{Yanfei Guo}
\orcid{0000-0002-3731-5423}
\anl{}
\author{Rajeev Thakur}
\orcid{0000-0002-5532-3048}
\anl{}
\ifacm
\acmYear{2023}
\acmConference[ICPP23]{International Conference on Parallel Processing}{August 07--13, 2023}{Salt Lake City, Utah}
\fi


\copyrightyear{2023}
\acmYear{2023}
\setcopyright{licensedusgovmixed}
\acmConference[ICPP 2023]{52nd International Conference on Parallel Processing}{August 7--10, 2023}{Salt Lake City, UT, USA}
\acmBooktitle{52nd International Conference on Parallel Processing (ICPP 2023), August 7--10, 2023, Salt Lake City, UT, USA}
\acmPrice{15.00}
\acmDOI{10.1145/3605573.3605599}
\acmISBN{979-8-4007-0843-5/23/08}

\begin{document}
\begin{abstract}
Partitioned communication was introduced in MPI 4.0 as a user-friendly interface to support pipelined communication patterns, particularly common in the context of MPI+threads.
It provides the user with the ability to divide a global buffer into smaller independent chunks, called 
\emph{partitions}, which can then be communicated independently.
In this work we first model the performance gain that can be expected when using partitioned communication.
Next, we describe the improvements we made to \mpich{} to enable those gains and provide a high-quality implementation of MPI partitioned communication.
We then evaluate partitioned communication in various common use cases and assess the performance in comparison with other MPI point-to-point and one-sided approaches.
Specifically, we first investigate two scenarios commonly encountered for small partition sizes in a multithreaded environment: thread contention and overhead of using many partitions. We  propose two solutions to alleviate the measured penalty and demonstrate their use.
We then focus on large messages and the gain obtained when exploiting the delay resulting from computations or load imbalance. 
We conclude with our perspectives on the benefits of partitioned communication and the various results obtained.
\end{abstract}

\maketitle

\pagestyle{plain}
\thispagestyle{plain}

\section{Introduction}
\label{sec:intro}

A hybrid MPI+threads model is commonly used nowadays to program parallel systems comprising nodes with multiple cores or accelerators such as GPUs. A common scenario in such a model is that multiple threads perform operations at different locations on the same buffer.
In this situation, a bulk synchronization of the threads followed by a single communication is usually the chosen approach to avoid heavy congestion on the MPI resources~\cite{Zambre:2022}, as illustrated in \Fref{fig_bulk_sync}.
However, the load imbalance between threads or the computational load can lead to some threads idling before the communication. It delays the start of the send operation and misses the opportunity to overlap communication with computation.
An alternative approach is the pipelined communication model, as illustrated in \Fref{fig_pipelined}.
Instead of sending the entire buffer using one thread, each thread now sends its own section of the buffer: each thread performs computations independently and communicates the results immediately.
Therefore, the first thread to end its computation gets a head start in the communication (also called the \eb{} effect).
While it enables the send operation to be started as soon as one thread completes the computation, it also brings another challenge of coordinating the communication from multiple threads.
This multithreaded MPI communication pattern usually scales poorly because of the contention for shared resources such as message queues and communication contexts.
Several approaches have been proposed to tackle this issue and provide performance for the pipelined communication pattern.
Some of the most well-known ones are  scalable endpoints~\cite{Zambre:2018}, finepoint communication~\cite{Grant:2019}, thread-based MPI implementation ~\cite{Kamal:2010}, and more recently \code{MPIX\_Stream}~\cite{Zhou:2022}.
Inspired by all these works, the MPI 4.0 standard~\cite{Message-Passing-Interface-Forum:2021} introduced point-to-point partitioned communication~\cite{Dosanjh:2021} to provide better support for the pipelined communication model and improve its adoption among users.
This new feature divides the communication buffer into non-overlapping \emph{partitions} where threads can operate individually. When one partition is ready, the thread marks it as ``ready to be sent''.
The send operation is completed once the main thread completes the communication, after all the partitions have been marked as ready.

This design allows the early threads to start the communication of partitions when they become ready.
It also offers two advantages compared with other MPI functionalities:
\begin{enumerate}
\item \textit{Easy-to-use multithreaded MPI communication.} The semantics of partitioned communication provides a simple interface to the user, hiding the complexity of multithreaded performance. It is now  easy to benefit from the \eb{} effect and achieve performance gain.
\item \textit{Flexibility in message transmission.} The new API gives the implementation the opportunity to perform optimizations in order to reduce the latency, otherwise  tedious to implement for users. A commonly considered approach is to aggregate partitions together into a single message in order to avoid the overhead for small partitions.
\end{enumerate}

\begin{figure*}[ht!]
\begin{minipage}[t]{0.49\textwidth}
\centering
\tikzstyle{branch}=[draw,fill=black,thick]
\tikzstyle{box}=[fill=black!2,rounded rectangle,inner sep=4pt,thick]
\tikzstyle{rbox}=[fill=black!2,rectangle,inner sep=4pt,thick]

\begin{tikzpicture}[>=latex']

\def\dy{0.8}

\draw[thick,black,->] (\dy/4,0) -- (10.5*\dy,0);
\node[] at (9.5*\dy,-\dy/2) (th0) {time};

\draw[thick,black] (\dy,-\dy) -- (\dy,4*\dy);

\node[] at (\dy/2,1*\dy/2) (th0) {\code{T0}};
\node[] at (\dy/2,3*\dy/2) (th0) {\code{T1}};
\node[] at (\dy/2,5*\dy/2) (th0) {\code{T2}};
\node[] at (\dy/2,7*\dy/2) (th0) {\code{T3}};

\def\start{6/4*\dy}

\draw[thin,black] (\start+2*\dy,0) -- (\start+2*\dy,4*\dy);
\draw[thin,black] (\start+3*\dy,0) -- (\start+3*\dy,4*\dy);
\draw[thick,black,<->] (\start+2*\dy,3.9*\dy) -- (\start+3*\dy,3.9*\dy) node[pos=0.5,anchor=south] {idle};

\draw[thin,black] (\start+9*\dy,0*\dy) -- (\start+9*\dy,4*\dy);
\draw[thick,black,<->] (\start+3*\dy,3.9*\dy) -- (\start+9*\dy,3.9*\dy) node[pos=0.5,anchor=south] {communication time};

\def\len{2.5*\dy}
\def\ty{1/4*\dy}
\def\xstart{\start}
\def\clen{2*\dy}
\draw[rbox,fill=asparagus,draw=black,thin] (\xstart,\ty) rectangle (\xstart+\clen,\ty + \dy/2) node[pos=0.5] {compute \#0};

\def\ty{5/4*\dy}
\def\xstart{\start}
\def\clen{3*\dy}
\draw[rbox,fill=asparagus,draw=black,thin] (\xstart,\ty) rectangle (\xstart+\clen,\ty + \dy/2) node[pos=0.5] {compute \#1};

\def\ty{9/4*\dy}
\def\xstart{\start}
\def\clen{2.75*\dy}
\draw[rbox,fill=asparagus,draw=black,thin] (\xstart,\ty) rectangle (\xstart+\clen,\ty + \dy/2) node[pos=0.5] {compute \#2};

\def\ty{13/4*\dy}
\def\xstart{\start}
\def\clen{2.25*\dy}
\draw[rbox,fill=asparagus,draw=black,thin] (\xstart,\ty) rectangle (\xstart+\clen,\ty + \dy/2) node[pos=0.5] {compute \#3};

\def\ty{1/4*\dy}
\def\clen{1.5*\dy}
\def\xstart{\start+3*\dy}
\def\xstart{\start+3*\dy}
\draw[rbox,fill=cadmiumorange,draw=black,thin] (\xstart,\ty) rectangle (\xstart+4*\clen,\ty + \dy/2) node[pos=0.5] {send [\#0 .\,\!.\,\!. \#3]};

\draw[ultra thick,black,firebrick] (\start,0) -- (\start,4*\dy) node[pos=.5,anchor=south,rotate=90] {\code{barrier}};
\def\clen{3.0*\dy}
\draw[ultra thick,black,firebrick] (\start+\clen,0) -- (\start+\clen,4*\dy) node[pos=.5,anchor=north,rotate=90] {\code{barrier}};

\end{tikzpicture}
\caption{Bulk thread synchronization followed by the send operation. The time idle due to imbalance and computation delays is wasted.}
\label{fig_bulk_sync}
\end{minipage}%
\hfill%
\begin{minipage}[t]{0.49\textwidth}
\centering
\tikzstyle{branch}=[draw,fill=black,thick]
\tikzstyle{box}=[fill=black!2,rounded rectangle,inner sep=4pt,thick]
\tikzstyle{rbox}=[fill=black!2,rectangle,inner sep=4pt,thick]

\begin{tikzpicture}[>=latex']

\def\dy{0.8}

\draw[thick,black,->] (\dy/4,0) -- (10.5*\dy,0);
\node[] at (9.5*\dy,-\dy/2) (th0) {time};

\draw[thick,black] (\dy,-\dy) -- (\dy,4*\dy);

\node[] at (\dy/2,1*\dy/2) (th0) {\code{T0}};
\node[] at (\dy/2,3*\dy/2) (th0) {\code{T1}};
\node[] at (\dy/2,5*\dy/2) (th0) {\code{T2}};
\node[] at (\dy/2,7*\dy/2) (th0) {\code{T3}};

\def\start{6/4*\dy}

\draw[thin,black] (\start+2*\dy,0) -- (\start+2*\dy,4*\dy);
\draw[thin,black] (\start+3*\dy,0) -- (\start+3*\dy,4*\dy);
\draw[thick,black,<->] (\start+2*\dy,3.9*\dy) -- (\start+3*\dy,3.9*\dy) node[pos=1.0,anchor=south east] {\eb{} gain};

\draw[thin,black] (\start+8*\dy,0*\dy) -- (\start+8*\dy,4*\dy);
\draw[thick,black,<->] (\start+3*\dy,3.9*\dy) -- (\start+8*\dy,3.9*\dy) node[pos=0.5,anchor=south] {communication time};

\def\len{2.5*\dy}
\def\ty{1/4*\dy}
\def\xstart{\start}
\def\clen{2*\dy}
\draw[rbox,fill=asparagus,draw=black,thin] (\xstart,\ty) rectangle (\xstart+\clen,\ty + \dy/2) node[pos=0.5] {compute \#0};
\def\slen{1.5*\dy}
\def\sstart{\xstart+\clen}
\draw[rbox,fill=cadmiumorange,draw=black,thin] (\sstart,\ty) rectangle (\sstart+\slen,\ty + \dy/2) node[pos=0.5] {send \#0};
\draw[thin,black,dashed] (\sstart+\slen,0) -- (\sstart+\slen,4*\dy);

\def\ty{5/4*\dy}
\def\xstart{\start}
\def\clen{3*\dy}
\draw[rbox,fill=asparagus,draw=black,thin] (\xstart,\ty) rectangle (\xstart+\clen,\ty + \dy/2) node[pos=0.5] {compute \#1};
\def\slen{1.5*\dy}
\def\sstart{\xstart+6.5*\dy}
\draw[rbox,fill=cadmiumorange,draw=black,thin] (\sstart,\ty) rectangle (\sstart+\slen,\ty + \dy/2) node[pos=0.5] {send \#1};

\def\ty{9/4*\dy}
\def\xstart{\start}
\def\clen{2.75*\dy}
\draw[rbox,fill=asparagus,draw=black,thin] (\xstart,\ty) rectangle (\xstart+\clen,\ty + \dy/2) node[pos=0.5] {compute \#2};
\def\slen{1.5*\dy}
\def\sstart{\xstart+5*\dy}
\draw[rbox,fill=cadmiumorange,draw=black,thin] (\sstart,\ty) rectangle (\sstart+\slen,\ty + \dy/2) node[pos=0.5] {send \#2};
\draw[thin,black,dashed] (\sstart+\slen,0) -- (\sstart+\slen,4*\dy);

\def\ty{13/4*\dy}
\def\xstart{\start}
\def\clen{2.25*\dy}
\draw[rbox,fill=asparagus,draw=black,thin] (\xstart,\ty) rectangle (\xstart+\clen,\ty + \dy/2) node[pos=0.5] {compute \#3};
\def\slen{1.5*\dy}
\def\sstart{\xstart+3.5*\dy}
\draw[rbox,fill=cadmiumorange,draw=black,thin] (\sstart,\ty) rectangle (\sstart+\slen,\ty + \dy/2) node[pos=0.5] {send \#3};
\draw[thin,black,dashed] (\sstart+\slen,0) -- (\sstart+\slen,4*\dy);

\def\ty{1/4*\dy}

\draw[ultra thick,black,firebrick] (\start,0) -- (\start,4*\dy) node[pos=.5,anchor=south,rotate=90] {\code{barrier}};
\def\sstart{\xstart+6.5*\dy}
\draw[ultra thick,black,firebrick] (\sstart+\slen,0) -- (\sstart+\slen,4*\dy) node[pos=.5,anchor=north,rotate=90] {\code{barrier}};

\end{tikzpicture}
\caption{Pipelined send operations, initiated from each thread. The imbalance and computation delays provide a gain through the \eb{} effect.}
\label{fig_pipelined}
\end{minipage}
\end{figure*}

\rev{Partitioned communication has} been supported in \mpich{} since the release of MPI 4.0. However, the initial implementation is focused primarily on correctness instead of performance.
In this work we present improvements to the implementation, enabling the user to achieve the expected performance gains.
Specifically, we are now able to
(1) aggregate small partitions together; (2) if required by the user, reduce thread congestion when performing communication; and (3) reduce the time-to-solution using the \eb{} effect.
In \Sref{sec:validation} we present some background and assess the expected gain of using the pipelined communication pattern for very small and very large messages.
In \Sref{sec:implementation} we present the improvements to the existing implementation in \rev{\mpich{}}.
This work puts a particular focus on the user experience and measurable gain.
To validate and assess the obtained performance, in \Sref{sec:results} we compare the improved implementation in \mpich{} with other MPI 3.1 approaches, relying on both the point-to-point and one-sided semantics.
In \Sref{sec:conclusion} we present our conclusions  and  discuss future directions.

\subsection*{Related work and novelty}
\rev{Prior contributions have focused on assessing  the partitioned communication benefits.
In \cite{Dosanjh:2021}, the authors present initial performance metrics, with an emphasis on the perceived bandwidth metric for large messages.
More recently, in \cite{Hassan-Temucin:2023} the authors use four metrics to measure the performance: overhead, perceived bandwidth, application availability, and \eb{} communication.
Still with a focus on large messages, they  describe the behavior with different noise models and detail the usage for sweep- and halo-based algorithms.
Partitioned collective communication has also been proposed in \cite{Holmes:2021} as an extension to the MPI 4.0 semantics.
Despite those efforts, a comprehensive evaluation of the newly proposed semantics against existing ones is still missing, especially for small message sizes.
With this work we aim to bridge that gap and therefore guide MPI users in making informed and evidence-driven choices for their own applications.
}

\section{Pipelined communication: Performance model and implementation approaches}
\label{sec:validation}

A detailed view of pipelined communication is presented in \Fref{fig_benchmark}, where we  highlight the different operations performed by each thread.
Our presentation is intentionally  general:  we will later detail how to implement the pattern using different strategies.
To initiate the pattern, the master thread performs a \code{start} operation, which implies a thread barrier afterward.
Then each thread performs computations and marks the partition as \code{ready}.
The master thread can finalize the communication using \code{wait}, which usually entails a thread barrier beforehand depending on the MPI API used.

\subsection{Performance measurements}
Measuring the performance of the pipelined communication pattern can be done in different ways \cite{Dosanjh:2016,Hassan-Temucin:2023}.
In our case we focus on the user experience, and therefore the time-to-solution is the most relevant metric.
As illustrated in \Fref{fig_benchmark} in \textcolor{firebrick}{red}, the latter runs from the \code{start} operation up to the completion of the communication on the receiver side.
Since we benchmark the MPI-related operations and not the computation, we remove the time of each thread spent in the computation.
By doing so we have a measure relevant to the user, namely, the overhead coming from the communications only.
We note that this metric is close to the perceived bandwidth proposed by \cite{Dosanjh:2021}. However, we go a step further and include the \code{start} operation and the following thread barrier into our metric.

\subsection{Performance prediction}
\label{sec_perf_prediction}
The performance of the pipelined communication pattern compared with the bulk thread-synchronization can be expressed as 
\be
\eta = \dfrac{T_b}{T_{p}}
\eec
where $T_b$ is the communication time with the bulk thread synchronization and $T_{p}$ is the communication time with the pipelined communication pattern.
For large message sizes, the value of $\eta$ is driven by the delay coming from the computations and the load imbalance between threads.
For small message sizes, however, the latency of the communication will prevail over the delay time and dictate the gain.
In this section we  further describe these two factors and the performance gain that a user might expect in both situations.

\subsubsection{Large messages and delay time}
\label{sec_perf_prediction_large}
For very large messages, the communication time will  be given by $\spart/\beta$, where $\beta$ is the bandwidth of the network and $\spart$ is the size of one partition.
The communication time associated with  bulk thread synchronization is given by
\be
T_b \approx \npart \; \dfrac{\spart}{\beta}
\eec
where a total of $\npart$ very large partitions will be used.
\rev{Introducing $\theta$ as the number of partitions per thread and $N$ the number of threads, we obtain that $\npart = N \theta$.}
When using pipelined communication, the communication time, $T_{p}$, is given by
\be
T_{p} \approx \max \left\{(\npart-1) \; \dfrac{\spart}{\beta}  - D \; ,\; 0\right\} +  \dfrac{\spart}{\beta}
\eec
where $D$ is the delay time between the first and the last partition to be ready and the $\max$ ensures that we overlap at most the communication time of the $\npart-1$ first partitions with the delay.
The latter is assumed to depend linearly on the partition size, leading to the definition of the delay rate, $\gamma$, such that $D = \gamma\; \spart $.
The expression of $\gamma$ is itself a function of $\theta$ and other parameters; see \Sref{sec_delay_rate}.

Combining the equations, we obtain the theoretical gain associated with the pipelined communication as
\be
\eta = \dfrac{T_{b}}{T_p} = \dfrac{N \theta}{ \max \left\{ N \theta - \gamma_{\theta} \; \beta \;,\; 1\right\}}
\eed
In practice, the bandwidth $\beta$, the algorithm, and the number of threads $N$ are usually fixed.
The gain is then a function of the number of partitions per thread, $\theta$, a user-controlled parameter.
For example, with $\theta = 1$, $\beta=25$\GBs{}, and $N=8$ threads, typical values are $\gamma \approx \left[ 1 \; ; \; 10 \right]$ \mus{}\per{}\MB{}, which lead to $\eta = 1.003$ and $\eta = 1.032$, respectively.
However, increasing the number of partitions per thread would enable the communication to be started earlier and therefore increase the delay rate.
With $\theta = 8$, the value of $\gamma$ goes up to $\approx 1000$ \mus{}\per{}\MB{} and the gain to $\eta = 1.641$, leading to a more significant benefit.

We conclude that using multiple partitions per thread is therefore crucial for performance.
However, it is hard to achieve in practice at large message sizes  because $\theta$ is inversely proportional to the size of the message, and our assumption of nonsignificant latency quickly becomes invalid.

\subsubsection{Small messages and latency}
\label{sec_perf_prediction_small}
In some situations the latency overwhelms the cost of the communication, due either  to a small buffer or to a very large number of partitions per thread.
In practice, this is usually the case for messages $\ll16$\kB{}. %
Furthermore, the delay generated by the computations is irrelevant for those small messages.
Assuming a delay rate of $\gamma=100$\mus{}\per{}\MB{} (see the preceding section) and a latency of $1$ \mus{}, a buffer of $1$\kB{} would generate enough delay to offset \rev{$10\%$} of the latency of a single message.

Therefore, assuming that latency is the only relevant metric and that the delay is negligible, we  obtain
\be
\eta = \dfrac{1}{N \theta}
\eed
In this situation, issuing multiple messages in the pipeline communication scheme will increase  the overhead and decrease the performance.
To avoid this issue, the user must either aggregate messages or decrease the number of partitions (and therefore increase the partition size).
Furthermore, this prediction does not take into account the thread contention that will also impact the performance for \rev{small messages}.

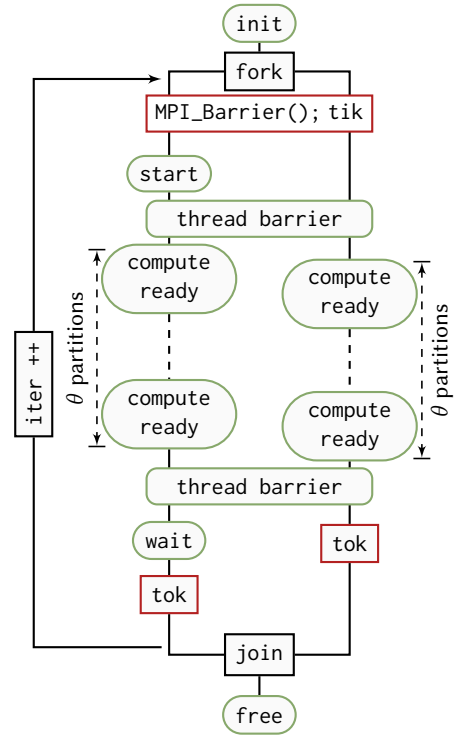
\begin{figure}[ht!]
\centering
\tikzstyle{branch}=[draw,fill=black,thick]
\tikzstyle{box}=[fill=black!2,rounded rectangle,inner sep=4pt,thick]
\tikzstyle{rbox}=[fill=black!2,rectangle,inner sep=4pt,thick]

\begin{tikzpicture}[>=latex']

\def\dy{0.8}
\def\middlesend{-4.0*\dy}
\def\middlerecv{2.0*\dy}
\def\middle{-\middlesend}
\def\yshift{1.5*\dy}
\def\thRsend{\middlesend+1.0*\yshift}
\def\thLsend{\middlesend-1.0*\yshift}
\def\thRrecv{\middlerecv+1.0*\yshift}
\def\thLrecv{\middlerecv-1.0*\yshift}

\def\yy{-0.5*\dy}
\node[box,draw=asparagus] at (\middlesend,\yy) (init) {\code{init}};
\def\yy{-1.3*\dy}
\node[rbox,draw=black] at (\middlesend,\yy) (fork){\code{fork}};
\draw[branch,-] (init) -- (fork);
\def\yy{-2.0*\dy}
\node[rbox,draw=firebrick] at (\thRsend,\yy) (tik_th1) {};
\node[rbox,draw=firebrick] at (\thLsend,\yy) (tik_th2) {};
\draw[branch,-,fill=none] (fork) -| (tik_th1);
\draw[branch,-,fill=none] (fork) -| (tik_th2);
\def\yy{-3.0*\dy}    
\node[box,draw=asparagus] at (\thLsend,\yy) (th2_start) {\code{start}};
\draw[branch,-] (tik_th2) -- (th2_start);
\def\yy{-4.0*\dy}
\node[box,draw=asparagus,text width=1.25cm,align=center] at (\thLsend,\yy-0.75*\dy) (th2_sleep) {\code{compute} \code{ready}};
\draw[branch,-] (th2_sleep) -- (th2_start);

\def\yy{-5.0*\dy}
\node[box,draw=asparagus,text width=1.25cm,align=center] at (\thRsend,\yy) (th1_sleep) {\code{compute} \code{ready}};

\draw[branch,-,fill=none] (th1_sleep) -- (tik_th1);

\def\yy{-6.0*\dy}
\node[box,draw=asparagus,text width=1.25cm,align=center] at (\thLsend,\yy-1.0*\dy) (th2_ready) {\code{compute} \code{ready}};
\draw[branch,-,fill=none,dashed] (th2_sleep) -- (th2_ready);

\def\yy{-7.0*\dy}
\node[box,draw=asparagus,text width=1.25cm,align=center] at (\thRsend,\yy-0.2*\dy) (th1_ready) {\code{compute} \code{ready}};
\draw[branch,-,fill=none,dashed] (th1_sleep) -- (th1_ready);

\def\nxl{\thLsend-0.8*\yshift}
\def\nxr{\thRsend+0.8*\yshift}
\draw [|<->|,semithick,dashed] (\nxr,-4.45*\dy) -- node[pos=0.5,anchor=north,rotate=90] {$\theta$ partitions}  (\nxr,-7.75*\dy);

\draw [|<->|,semithick,dashed] (\nxl,-4.25*\dy) -- node[pos=0.5,anchor=south,rotate=90] {$\theta$ partitions}  (\nxl,-7.55*\dy);

\def\yy{-9.0*\dy}
\node[box,draw=asparagus] at (\thLsend,\yy-0.1*\dy) (th2_wait) {\code{wait}};
\draw[branch,-,fill=none] (th2_ready) -- (th2_wait);
\node[rbox,draw=firebrick] at (\thRsend,\yy-0.15*\dy) (th1_tok) {\code{tok}};
\def\yy{-10.0*\dy}
\node[rbox,draw=firebrick] at (\thLsend,\yy-0*\dy) (th2_tok) {\code{tok}};
\draw[branch,-,fill=none] (th1_tok) -- (th1_ready);
\draw[branch,-,fill=none] (th2_tok) -- (th2_wait);

\def\yy{-11.0*\dy}
\node[rbox,draw=black] at (\middlesend,\yy) (parallel_end){\code{join}};
\draw[branch,-,fill=none] (parallel_end) -| (th2_tok);
\draw[branch,-,fill=none] (parallel_end) -| (th1_tok);

\def\yy{-11.0*\dy}
\node[rbox,draw=black,rotate=90] at (\middlesend-2.5*\yshift,\yy+4.5*\dy) (iter) {\code{iter ++}};
\draw[branch,-,fill=none] (\thLsend-0.1,\yy+0.1) -| (iter);
\draw[branch,->,fill=none] (iter) |- (\thLsend-0.1,-1.3*\dy-0.1);
\def\yy{-12.0*\dy}
\node[box,draw=asparagus] at (\middlesend,\yy) (free) {\code{free}};

\draw[branch,-] (parallel_end) -- (free);

\def\yy{-2*\dy}
\draw[rbox,draw=firebrick,fill=white] (\thRsend+0.25*\yshift,\yy-0.3*\dy) rectangle (\thLsend-0.25*\yshift,\yy+0.3*\dy) node[pos=.5] () {\code{MPI\_Barrier();}~\code{tik}};
\def\yy{-3.75*\dy}
\draw[box,draw=asparagus,rounded corners] (\thRsend+0.25*\yshift,\yy-0.3*\dy) rectangle (\thLsend-0.25*\yshift,\yy+0.3*\dy) node[pos=.5] () {\code{thread barrier}};
\def\yy{-7.5*\dy}
\draw[box,draw=asparagus,rounded corners] (\thRsend+0.25*\yshift,\yy-1.0*\dy) rectangle (\thLsend-0.25*\yshift,\yy-0.4*\dy) node[pos=.5] () {\code{thread barrier}};

\end{tikzpicture}
\caption{Benchmark for the pipelined communication pattern, illustrated with two threads. The application to each of the proposed implementation is summarized in \Cref{tab:benchmark-sender,tab:benchmark-receiver}.
Red boxes represent operations added for benchmarking purposes.}
\label{fig_benchmark}
\end{figure}

\subsection{Possible user approaches}
\label{sec_poss_user_impl}

To implement the pipelined communication pattern, the user might consider different approaches, divided into three categories:  partitioned communications (MPI 4.0), point-to-point (MPI 3.1), and RMA-based (MPI 3.1) APIs.
\rev{In \Cref{fig_benchmark} we present the general template for the different implementations.
For each of the steps, the list of MPI API calls actually used is summarized in \Tref{tab:benchmark-sender} for the sender side and in \Tref{tab:benchmark-receiver} for the receiver side.
The actual implementation for each of them is available at \cite{racetrack:2023}.}

\begin{table*}[t!]
\caption{MPI operations for the sender side.}
\label{tab:benchmark-sender}
\begin{tabularx}{\textwidth}{Z{3.5cm} | YYYY}
 & \code{init}                        & \code{start}          & \code{ready}        & \code{wait} \\
\toprule
\ptpart     & \code{MPI\_Psend\_init} & \code{MPI\_Start} &  \code{MPI\_Pready} & \code{MPI\_Wait} \\
\midrule
\ptsingle & \code{MPI\_Send\_init} &   &  & \code{MPI\_Start}\newline \code{MPI\_Wait} \\
\midrule
\ptmany & \code{MPI\_Comm\_dup} \newline \code{MPI\_Send\_init} &   &\code{MPI\_Start}\newline \code{MPI\_Wait}  &   \\
\midrule
\RMAsingle & \code{MPI\_Comm\_dup} \newline \code{MPI\_Win\_create}\newline \code{MPI\_Win\_lock}  & \code{MPI\_Recv} &\code{MPI\_Put} & \code{MPI\_Win\_flush} \newline \code{MPI\_Send}  \\
\midrule
\RMAmany & \code{MPI\_Win\_create}\newline \code{MPI\_Win\_lock}  & \code{MPI\_Recv} &\code{MPI\_Put} \newline \code{MPI\_Win\_flush} &  \code{MPI\_Send}  \\
\midrule
\RMAsingleactive & \code{MPI\_Comm\_dup} \newline \code{MPI\_Win\_create} & \code{MPI\_Start} &\code{MPI\_Put} & \code{MPI\_Complete}  \\
\midrule
\RMAactive &\code{MPI\_Win\_create} & & \code{MPI\_Start} \newline \code{MPI\_Put} \newline \code{MPI\_Complete} &    \\
\bottomrule
\end{tabularx}
\end{table*}

\begin{table*}[t!]
\caption{MPI operations for the receiver side.}
\label{tab:benchmark-receiver}
\begin{tabularx}{\textwidth}{Z{3.5cm} | YYYYYY}
 & \code{init}                        & \code{start}    & \code{ready}     & \code{wait} \\
\toprule
\ptpart     & \code{MPI\_Precv\_init} & \code{MPI\_Start}  & \code{MPI\_Parrived} & \code{MPI\_Wait} \\
\midrule
\ptsingle & \code{MPI\_Recv\_init} & \code{MPI\_Start}  &  & \code{MPI\_Wait}  \\
\midrule
\ptmany & \code{MPI\_Comm\_dup} \newline \code{MPI\_Recv\_init} &  &\code{MPI\_Start} \newline \code{MPI\_Wait} &  \\
\midrule
\RMAsingle \newline \RMAmany  &\code{MPI\_Win\_create} & \code{MPI\_Send}   & &  \code{MPI\_Recv}  \\
\midrule
\RMAsingleactive \newline \RMAactive & \code{MPI\_Win\_create} & \code{MPI\_Post}  &  & \code{MPI\_Wait}\\
\bottomrule
\end{tabularx}
\end{table*}
\subsubsection{Partitioned communication}
The partitioned communication API provides the user with a simple way of exploiting the pipeline communication pattern.
The communication is initialized during the call to \code{MPI\_Psend\_init}.
Then, the main thread calls \code{MPI\_Start} on the partitioned requests, which is followed by a thread barrier.
Once the computation on a partition is completed, the thread can  call \code{MPI\_Pready} to signal to MPI that the partition is ready to be sent.
After a barrier, the master thread  calls the \code{MPI\_Wait} function to complete the communication.
The receiver side is similar. \code{MPI\_Precv\_init} is used to initialize the communication, along with \code{MPI\_Start} to start an iteration.
A thread can then query the status of a partition using \code{MPI\_Parrived}, and the master thread completes the communication using \code{MPI\_Wait}.

\subsubsection{Point-to-point MPI 3.1}
\rev{A first approach is be to use a single message to communicate once the threads have completed their work on the different partitions.}
This strategy implements a bulk thread synchronization instead of the pipelined communication and is referred to as \ptsingle{}.
After a thread barrier, the master thread issues the persistent communication with \code{MPI\_Start}. The receiver uses a single persistent request to receive the message.
\rev{Another approach is to send one message from every thread as soon as the computation is over.}
This approach is denoted here as \ptmany{}.
To avoid competition on the same resource, we first duplicate the communicator per thread. Different communicators will be mapped to different communication contexts, hence removing the contention between the threads~\cite{Zambre:2020}.
Then each thread can send and receive its own partition independently of the status of other threads.
This approach, more complicated for the user than the partitioned communication, is the traditional way of taking advantage of a pipelined communication pattern.

\subsubsection{One-sided communication}
\label{sec_user_rma}
\rev{Our third category of implementation uses the MPI one-sided (RMA) semantics.}
Similar to the point-to-point variations, it can be implemented either on a single window shared by all threads or by using one window per thread, each over the entire buffer.
The approaches also can be distinguished by their synchronization API: active or passive.
By design, a send-receive operation is an active RMA communication pattern, which means that the target of the RMA call is involved in the communication.
This pattern can be naturally implemented with the active synchronization API. However, an enhanced use of the passive synchronization API can also be used to implement an active communication pattern, at the cost of added synchronization.

In the active synchronization API, the origin opens and closes the \emph{access epochs} through \code{MPI\_Start} and \code{MPI\_Complete}, respectively. The target controls when its memory is exposed (\emph{exposure epochs}) using \code{MPI\_Post} and \code{MPI\_Wait}.
The main advantage of the active API is to offer  explicit control to the user on the target readiness to handle the data.
The active synchronization approach on a single window and on many windows \rev{is denoted} as \RMAsingleactive{} and \RMAactive{}, respectively, in the rest of this work.

In contrast, the passive synchronization API is based only on managing the \emph{access epochs}.
One must still control the \emph{exposure epochs}, which can be done by using $0$\B{} send and receive messages.
We  note that ensuring progress with passive synchronization can be a challenge, especially when no global progress is done in \mpich{}. Different approaches exist to  address this issue;  see \cite{SI:2015} for a thoughtful discussion of them.
For our specific case, we have chosen to use \code{MPI\_MODE\_NOCHECK} when locking the window, to avoid requiring the receiver to be involved in the synchronization at that stage.
In the rest of this work, \rev{we denote} the passive synchronization on a single window and on multiple windows, respectively, as \RMAsingle{} and \RMAmany{}.

\section{Partitioned communication implementation in \mpich{}}
\label{sec:implementation}

In this section we briefly describe the existing implementation of the partitioned communications in \mpich{} and the improvements made as part of this work.
This section focuses on the underlying mechanisms, which are useful for understanding the performance obtained.

\subsection{Existing implementation}

The current implementation of partitioned communication in \mpich{} uses a single-message approach, done through the active messaging (AM) code path.
When the user calls \code{MPI\_Psend\_init}, an atomic counter is associated with the partition request. A ``ready-to-send" (RTS) message is sent to the receiver with some of the basic information about the size of the data and the number of partitions.
During \code{MPI\_Start}, the counter is set to the number of partitions given by the user, plus one. The ``plus one" takes into account that for each iteration the sender has to wait for  a ``clear-to-send" (CTS) message  from the receiver.
Because of the AM nature, this mandatory CTS avoids early sends from the sender to a receiver still in the previous iteration. Upon receiving the CTS, the sender will decrement the counter by one.
Then, when a partition is ready and the user calls \code{MPI\_Pready}, the counter is decremented by one.
Once all the partitions are ready and the CTS has been received, the value of the counter is zero, and the whole buffer is sent to the receiver.

The use of AM, together with the CTS needed at each iteration, delivers a semantically correct implementation, yet not the expected performance for the user.
Specifically we would see no benefit of the \eb{} effect coming with the pipelined communication.

\subsection{Improvements}
\label{sec_impl_improvements}
We have improved the implementation in \mpich{} to use multiple internal tag-matched messages, instead of a single AM communication.
Another option would have been to rely on an RMA-supported implementation, as suggested in \cite{Dosanjh:2021}.
We decided not to follow this strategy for two reasons.
First, the difference between the two approaches matters only for small messages.
Second, an RMA-based approach requires exposure control (see \Sref{sec_user_rma}), which increases the overhead. To alleviate this overhead, we rely on the repetitive use of a put operation, faster than a tag-matching send.
For small messages, however,  optimal performance is obtained with a few messages (see \Sref{sec_perf_prediction_small}). In this configuration an RMA-based implementation is then slower, as illustrated in \Sref{sec_res_improvement}.

\subsubsection{Initialization}
\label{sec_implementation_init}
During the initialization, the sender and the receiver will agree on using the tag-matching code path and on a fixed number of messages to be sent.
The tag matching can  be used only if there is enough tag space to isolate the traffic of partitioned communication from other communications coming from the user.
The sender keeps a count of the number of partitioned requests created for each of the receiver ranks. If that number exceeds the tag space reserved for the partitioned communications,  the AM code path is used instead.

\rev{
The sender and the receiver have to agree on the number of messages to be actually sent.
The most general protocol is to let the receiver decide this number once the RTS has been received. Then, that information is sent to the sender with the CTS.
However, this general approach incurs a performance overhead for two reasons.
First, the sender has to wait for the CTS during the first iteration.
Second, a CTS does not naturally fit within a tag-matching communication protocol.
Another approach would be to let the sender decide on the number of messages to be sent.
However, this strategy adds complexity when the sender and/or the receiver uses noncontiguous datatypes.
In this case the receiver might receive a partial datatype, and dealing with this scenario efficiently is challenging.
}

\rev{
In this work we have chosen to implement the first, yet suboptimal, approach.
The receiver decides the number of messages to be sent using $\mygcd{\npart^{\text{send}}}{\npart^{\text{recv}}}$, which guarantees that a partition will contribute only to a single message.
In our implementation, the receiver is also in charge of message aggregation, based on the user-defined threshold \code{MPIR\_CVAR\_PART\_AGGR\_SIZE}. %
This value is used as an upper bound for aggregation: if the size of multiple messages fit within the prescribed threshold, then they are aggregated together.
The number of messages obtained from this logic is then sent to the sender as part of the CTS.
}

We note that the use of \code{MPI\_Parrived} is in contradiction with message aggregation.
Indeed, the former is used to reduce the overhead by exploiting a coarser-grained communication strategy, whereas  the latter suggests that the user could exploit a fine-grained communication pattern.
In our implementation we have chosen to optimize \mpich{} toward achieving low latency, and therefore we have not spent much effort in optimizing the usage of \code{MPI\_Parrived}.

\subsubsection{Sending and receiving partitions}
\label{sec_impl_improvements_send}
On the sender side, at each iteration, each message to be sent is associated with an atomic counter whose value is set to the number of partitions contributing to the message.
When a partition is marked as ready by the user, the associated counter is decremented. If the value reaches zero, the message is then sent using tag-matching or an AM send/receive \mpich{} internal API.
On the receiver side, each message is associated with a receive request. The user can then query the reception of a given partition, hence reading the status of the request.

We also allow the user to use different communication resources (known as \vci{} in \mpich{}) to send different partitions. 
This is done by encoding the source and destination \vci{} id into the tag, using an experimental feature in \mpich{}.
With this, we reduce the thread congestion that occurs when sending from multiple threads using the same resource.
However, despite the ubiquitous multithreaded context when using partitioned communication semantics, the user has no standard way of conveying the thread granularity to the MPI implementation.
Therefore, our implementation assumes a round-robin attribution of the threads to the partitions.
This assumption is inflexible and likely to break when used in practice with $\theta > 1$.
We note that info hints provided during communication initialization or the usage of \code{MPIX\_Stream}~\cite{Zhou:2022} with partitioned communication could be used to express the thread granularity to MPI.
Such improvements are left for future work.

In summary, the improvements offer better performance than the existing implementation, as detailed in \Sref{sec_res_improvement}. Moreover, the user now has the opportunity to take advantage of three possible gains:
\begin{enumerate}
\item \textit{Thread Congestion} (experimental): When multiple threads send different partitions simultaneously, they will compete for the same resources. The congestion is overwhelming especially for small partitions. To alleviate this issue, we use an experimental \mpich{} capability to use different resources for each partition. Further details can be found in \Sref{sec_res_vci}.
\item \textit{Message Aggregation}: When dealing with small partitions sizes, different partitions can be aggregated together under a single message to reduce the overhead. Results on this performance gain are detailed in \Sref{sec_small_part}.
\item \textit{Early-Bird Effect}: When sending large partitions, the user now benefits from the gain offered by the pipelined communication model as shown in \Sref{sec_eb_effect} .
\end{enumerate}

\newcommand{\legfull}[1]{\ptsingle \CaptionSingle, \ptmany \CaptionMulti, \ptpart \CaptionPart, \RMAactive \CaptionRMAActive, \RMAsingleactive \CaptionRMASingleActive, \RMAmany \CaptionRMAMulti, \RMAsingle \CaptionRMASingle}
\newcommand{\legnoactive}[1]{\ptsingle \CaptionSingle, \ptmany \CaptionMulti, \ptpart \CaptionPart, \RMAmany \CaptionRMAMulti, \RMAsingle \CaptionRMASingle}
\newcommand{\legptp}[1]{\ptsingle \CaptionSingle, \ptmany \CaptionMulti, \ptpart \CaptionPart}

\section{Performance Results}
\label{sec:results}

\rev{
In this section we compare the performance for each of the possible pipelined communication implementations (see \Cref{sec_poss_user_impl}), using the benchmark template described in \Cref{fig_benchmark}.
We first assess the benefits of the improved implementation and compare it with other MPI-3.1 approaches.
To streamline our analysis and avoid implementation artifacts, we assume that the number of partitions is the same on both the sender and the receiver side.
Next, we investigate the performance for small messages, together with the impact of message aggregation and thread congestion.
We then compare the different approaches to achieve pipelined communication with the expected gain when using large messages, as detailed in \Sref{sec_perf_prediction_large}.
}

All the results in this work have been obtained using \mpich{} and \code{ucx-1.13.1} between two nodes on MeluXina.\footnote{\rev{System in Luxembourg (\#379 top500 06/2023), with 73,344 cores (AMD EPYC 7H12 64C) on its CPU partition connected through a Mellanox $200$\emph{Gb}\per{}\ssec{} and $1.22$\mus{} latency HDR200-IB network}}
The \openmp{} threads are closely bound to the cores,\footnote{with \code{OMP\_PROC\_BIND=CLOSE} and \code{OMP\_PLACES=cores}} and the MPI processes are bound to as many cores as there are threads.\footnote{using \code{-bind-to cores:\$\{OMP\_NUM\_THREADS\}}}
The benchmark \cite{racetrack:2023} has been run for $150$ iterations and $1$ warm-up iteration to get rid of the overhead, explained in \Sref{sec_impl_improvements}. For each of the data results, we present the time as the average on the iterations (excluding the warm-up), and we  obtain a $90\%$ confidence interval assuming a Student's t-distribution.
To avoid network noise, we rerun the measure if the half-width of the confidence interval is larger than $5\%$ of the average time, with a maximum retry at $50$.
Confidence intervals are displayed as a shaded area around the results on figures displaying time.

\subsection{Improvement over existing implementation}
\label{sec_res_improvement}

\begin{figure}[ht!]
\pgfplotsset{
	legend columns=2,
	legend image code/.code={
        \draw[very thick,
        mark repeat=2,
        mark phase=2,
        dash phase=0pt,
        ] plot coordinates {(0pt,0pt) (15pt,0pt)};%
	}
}
\def\ith#1{1}
\def\vci#1{1}
\def\aggr#1{0}
\def\sleep#1{0}
\centering
\input{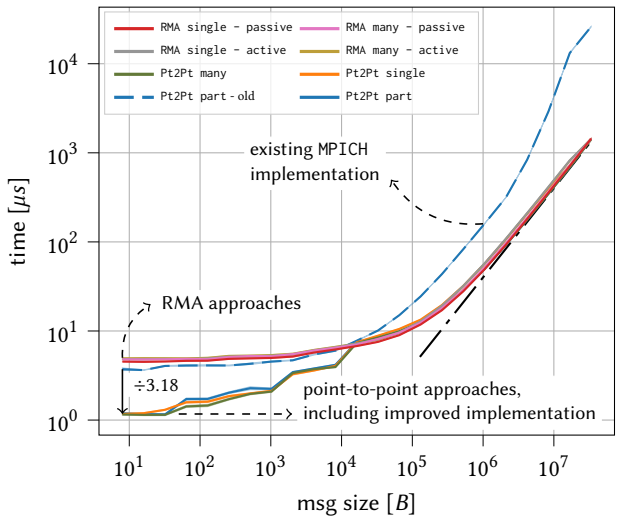}
\caption{Time across message sizes with \ith{} thread and \ith{} partition: \rev{comparison of the existing and improved partitioned communication implementation with other MPI-3.1 approaches.}}
\label{fig_time_1part_1thread_0sleep}
\end{figure}

To  demonstrate the gain of not using the AM path, we measure the time in the case of $N = 1$ threads, $\theta = 1$ partition per thread with no delay ($\gamma = 0$). Although not representative of the usual operation space of pipelined communications, the configuration is well suited to highlight the performance gain made possible by our improvements.
In \Fref{fig_time_1part_1thread_0sleep} we show the time needed by each of the approaches to complete the communication.
For reference, we also give the time corresponding to the theoretical bandwidth of the system ($25$\GBs).
The difference between the existing AM-based implementation \rev{(\ptpart{} - old)} and our improved version \rev{(\ptpart{})}  is noticeable for all message sizes. The latency associated with the copy needed in the AM code path implies a significant overhead and degrades the performance.
With the new implementation we  match the performance of the \ptsingle{} approach, as expected.

For point-to-point-based approaches, we note that the time jumps when switching protocols over the different messages sizes.
\rev{In particular, we note the change from the \code{short} protocol to the \code{bcopy} one between 1,024 and 2,048 \B~and to the \emph{rendez-vous} (\code{zcopy}) protocol from 8,192 to 16,384 \B~\cite{Shamis:2015}.}
The difference between the two families of approaches (point-to-point and RMA) is also clearly observed at small message sizes.
The RMA-based approaches require two additional synchronizations to be performed,  resulting in a larger overhead.
\rev{We also note that the gap vanishes when considering message sizes above the \emph{rendezvous} threshold.
The reason is that the bandwidth is dominant for large message sizes and all the approaches use the same communication protocol. The \code{zcopy} protocol used in point-to-point is actually based on the RMA network capability.}

\subsection{Small messages}

\subsubsection{Thread congestion}
\label{sec_res_vci}
\begin{figure}[ht!]
\pgfplotsset{legend columns=2}
\def\ith#1{32}
\def\vci#1{1}
\def\aggr#1{0}
\def\sleep#1{0}
\centering
\input{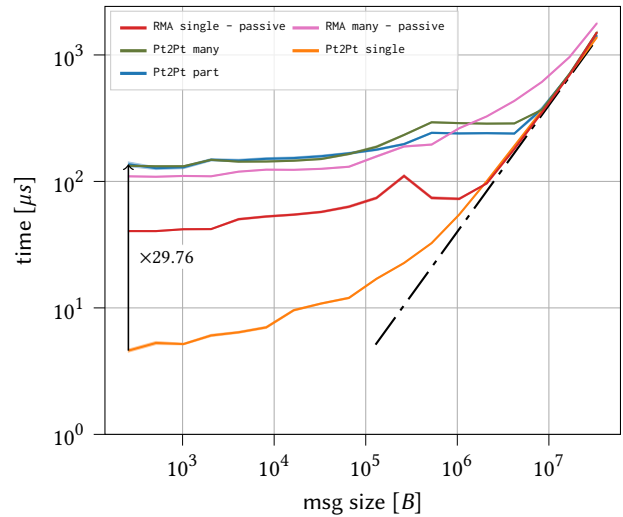}
\caption{Thread congestion: communication time across message sizes for \ith{} partitions with \ith{} threads.}
\label{fig_time_congestion_1vci}
\end{figure}

In practice, partitioned communication is used in a multithreaded environment, which will lead to thread congestion.
To highlight the issue, we present in \Fref{fig_time_congestion_1vci} the time needed to communicate when using $32$ threads and one partition per thread ($\theta = 1$).

As expected for small messages, the \ptsingle{} approach performs the best. The single message does not suffer from any of the downsides of the multithreading since the communication happens on one thread only.
However, we still note a higher latency compared with \Fref{fig_time_1part_1thread_0sleep} due to the needed synchronization barrier.
The \ptpart{} and \ptmany{} communication strategies both suffer from  thread contention, with little difference between the achieved overheads.
With the RMA-based passive synchronization approaches, the results are more sparse.
We observe that the RMA approaches using many windows (one per partition) suffer from an additional overhead compared with the single RMA window.
Regarding the \code{MPI\_Put} operations, there is no significant difference because in both cases the threads will compete for the same resource.
However, the \RMAmany{} approach adds an overhead in the progress engine compared with the \RMAsingle{} since it has to operate on multiple windows simultaneously.
This causes the upward shift observed in \Fref{fig_time_congestion_1vci}.

\begin{figure}[ht!]
\pgfplotsset{legend columns=2}
\def\ith#1{32}
\def\vci#1{32}
\def\aggr#1{0}
\def\sleep#1{0}
\centering
\input{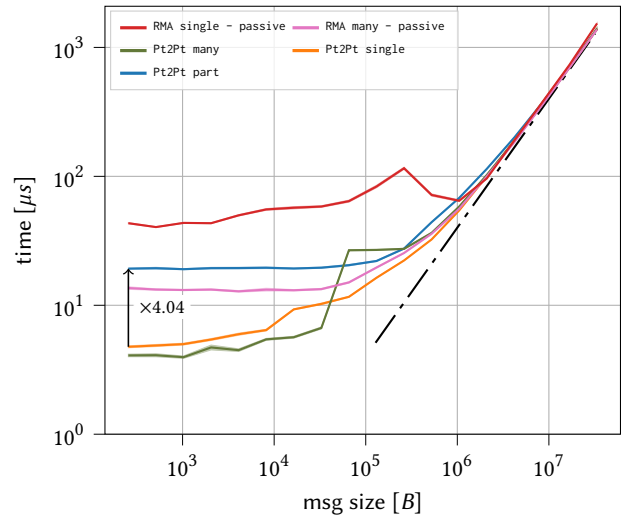}
\caption{Thread congestion: communication time across message sizes for \ith{} partitions with \ith{} threads using \vci{} VCIs.}
\label{fig_time_congestion_manyvci}
\end{figure}

In order to reduce the overhead, \mpich{} can be configured to use multiple \emph{virtual communication interfaces} (\vci{}s) \cite{Zambre:2020}.
This is achieved by using \code{MPIR\_CVAR\_NUM\_VCIS} to control the number of \vci{}s used by the implementation.
Different communicators/windows will then be mapped onto different \vci{}s, which allow multiple threads to access different resources.
As detailed in \Sref{sec_impl_improvements}, in the improved partitioned communication implementation, we map each partition on a different \vci{}s using a round-robin strategy.\footnote{if the user has used \code{--enable-vci-method=tag} during the configuration}
The results obtained when using one \vci{} per thread are presented in \Fref{fig_time_congestion_manyvci}.
In this setting, the \ptmany{} strategy reaches the same performance as the \ptsingle{} method. As we expect, different communicators are assigned to different \vci{}s, which then leads to no contention for the \ptmany{} approach. The \ptsingle{} method is further slightly penalized by the needed thread barrier before starting the send operation.
The \ptpart{} code path exploits the different \vci{}s as well; however, it still suffers from an overhead compared with \ptsingle{}. Compared with the non-\vci{} usage, we have decreased the cost of thread contention by a factor of $\approx 10$.
Regarding the RMA-based implementations, the \RMAmany{} is now faster than the \RMAsingle{}. The former approach relies on different \vci{}s (one per window) and therefore avoids the cost of contention.

As pointed out in \Sref{sec_impl_improvements_send}, the partitioned communication implementation may lack necessary information to avoid \vci{} contention in a multithreaded environment.
If this is the usage model of one's application, we would  favor the use of the \ptmany{} approach to get better performance.
In the rest of this work, we will consider a single \vci{} to illustrate the expected application context of the partitioned communication API.

\subsubsection{Partition aggregation}
\label{sec_small_part}

\begin{figure}[ht!]
 \pgfplotsset{legend columns=2}
\pgfplotsset{
	legend image code/.code={
		\draw[very thick,
			 mark repeat=2,
			 mark phase=2,
			 dash phase=0pt,
			 ] plot coordinates {(0pt,0pt) (20pt,0pt)};%
	}
}
\def\ith#1{4}
\def\ppt#1{32}
\def\vci#1{1}
\def\aggr#1{16384}
\def\sleep#1{0}
\centering
\input{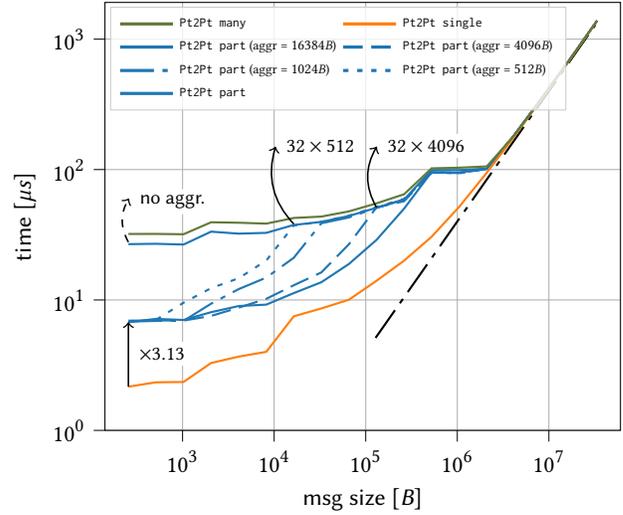}
\caption{Message aggregation: time for $\mathbf{\theta = \ppt{}}$ partitions per thread and \ith{} threads.}
\label{fig_time_aggr16384}
\end{figure}

To reduce the latency, one can also gather multiple partitions as a single message to avoid multiplication of the individual overheads.
As explained in \Sref{sec_implementation_init}, the user can use \mpich{'s} environment variable \code{MPIR\_CVAR\_PART\_AGGR\_SIZE} (in bytes) to request an upper bound on the aggregation size.
We note that this technique is compatible with other approaches such as \ptmany{}, but it would require significant code changes from the user.
\rev{To avoid interference with other delays and to focus our analysis on the message aggregation, we consider that all the partitions are ready immediately and processed in order by each thread.}
The results of this approach are shown in \Fref{fig_time_aggr16384}, where the messages are aggregated from 512 up to 16,384 \B{}.

We observe that the \ptpart{} reduces the overhead for small messages significantly compared with the \ptmany{} approach, which has the same performance as the no-aggregation based \ptpart{}.
\rev{For a given aggregation size, the number of messages actually sent increases with the size of the global buffer. Therefore, message aggregation is  beneficial for global message buffer size only below $\npart$ times the aggregation size.
As illustrated in \Fref{fig_time_aggr16384}, larger aggregation sizes will lead to a shift of that point toward large message sizes.}
Regardless of the size, however, we do not match the latency of the \ptsingle{} approach. The reason is the added overhead of partitioned communications such as the atomic update on the message counter performed by every partition when ready. The latter becomes more significant for an increased number of partitions.
\rev{For an infinite aggregation size and neglecting the overhead, \ptsingle{} is then the upper bound of performance.}

\subsubsection{Take-away}
For a user focused on a usage based on many small partitions, the partitioned communication offers an easy interface with little to no overhead compared with most advanced \ptp{} APIs.
Achieving the same performance using the standard MPI 3.1 API would complicate significantly the user's code, especially when message aggregation is desired.
\rev{For a user focused on performance, however, moving to a more explicit yet more complex API will take full advantage of the state-of-the-art features in \mpich{}.
For cases with many threads, we recommend the use of the \ptmany{} API with multiple VCIs. For cases with many partitions per threads and a few threads, we recommend instead the use of the \ptsingle{} approach to reduce the latency.}

\subsection{Large messages: benefit of the \eb{} effect}
\label{sec_eb_effect}
When using large messages, the delay generated by computations and load imbalance is significant, and it will drive the performance, as detailed in \Sref{sec_perf_prediction}.
One could measure the gain obtained for different values of $\theta$, different algorithm parameters, and so forth.
As demonstrated in \Sref{sec_delay_rate}, however, the value of $\gamma$  accurately models the delay obtained in those different situations.
Therefore, we rely instead on the value of $\gamma$ to characterize the delay obtained in various practical situations, which allows us to  measure only cases with $\theta = 1$.
Specifically, the last partition is delayed compared with the other $\npart-1$ partitions, where the delay time is given by $\gamma \spart$, where $\spart$ is the partition size.
Then, we measure the obtained bandwidth and compare it with the theoretical gain predicted in \Sref{sec_perf_prediction}.
In \Fref{fig_bw_sleep} we present the results with $\gamma = 100$ \mus\per{}\MB{}, which represents a practical delay; see \Sref{sec_delay_rate_numeric}.
With a total of $4$ partitions and $4$ threads, we measure a gain of $\approx 2.54$, close to the theoretical value of $2.67$.
The difference comes from the latency involved in the actual communications and the thread congestion, both left out of the model; see \Cref{sec_perf_prediction_large}.

\begin{figure}[ht!]
\def\ith#1{4}
\def\vci#1{1}
\def\aggr#1{0}
\def\sleep#1{100}
\centering
\input{figures/results/BW_sleep-\ith{}part-\ith{}thread-\sleep{}sleep-\vci{}vci-\aggr{}aggr.tex}
\caption{Gain obtained with the \eb{} effect ($\mathbf{\gamma = \sleep{}}$\mus\per{}\MB{}, which stands for a value of $\mathbf{\theta > 1}$) with 4 threads and 4 partitions}.%
\label{fig_bw_sleep}
\end{figure}

We  note that, as expected, the gain obtained from the \eb{} effect is independent of the approach used by the user.
Since the messages are dominated by bandwidth and have (almost) negligible latency, every possible variation of the MPI API will provide the same gain.
As  highlighted earlier, the partitioned communication API provides a  simple interface to the user in order to achieve that gain.
In real-life cases, however, the actual gain from an application perspective is tightly coupled to the size of the partitions and the delay achieved by the application.

In summary, the results of \Fref{fig_bw_sleep} represent perfectly the usage of partitioned communication for pipelined communication.
As expected by our performance modeling in \Sref{sec_perf_prediction}, we observe that with small messages it adds an overhead, due to the thread congestion and multiple latency costs. Therefore, for a fixed number of threads, using $\theta = 1$ will lead to the best performances.
For larger messages, however, the gain is significant as one hides communications behind computations. A larger number of partitions per thread ($\theta \gg 1$) will lead to a larger delay rate ($\gamma$) and therefore to a larger measured gain.
With this  example, we observe the trade-off to be around 100 \kB{}, a value driven by thread congestion.

\section{Conclusions}
\label{sec:conclusion}

In this work we investigate the pipelined communication pattern and the expected gain from the \eb{} effect.
First, we introduce a theoretical model to quantify the expected gain and to identify in which cases it will be beneficial.
Then, we present the improvements made to the \mpich{} implementation in order to deliver the expected performance to the user. %
Specifically, we provide three features: (1) thread congestion alleviation, (2) message aggregation, and (3) \eb{} effect gain obtained by starting the communication as soon as the data is ready.
Further, we explore various other approaches that rely on MPI 3.1  features and  could also be used to implement pipelined communication.
We study the use of point-to-point-based approaches, such as using a single message or one message per partition. We also consider various one-sided strategies relying on a single window, multiple windows, and active and passive synchronization.
We use a pipelined communication benchmark to compare them with the partitioned communication semantics, including the existing and improved implementations in \mpich{}.
Then, we investigate three specific cases across the spectrum of typical usage of the pipelined pattern.
First, we consider the case of small messages where multiple threads contend for the same MPI resources.
Relying on existing thread contention alleviation strategies in \mpich{}, we are able to reduce dramatically the associated overhead: Compared with a single-message approach, we reduce the penalty from a factor of $\approx 30$ to $\approx 4$.
Second, we demonstrate message aggregation and how it reduces the overhead associated with multiple messages to a single-message latency, at the cost of a few atomic updates.
With that strategy we are able to reduce the penalty factor from $\approx 10$ to $\approx 3$ compared with a single-message approach.
Third, we demonstrate how the user can benefit from a significant bandwidth improvement when using pipelined communication with large messages, even with thread contention.
In the context of the presented results, we measure a benefit for messages larger than $\approx 100$ \kB.
We also demonstrate that this benefit is agnostic to the type of method used (point-to-point or one-sided).
We conclude that the best configuration to use partitioned communication depends on the partition size. To avoid significant overhead with small messages, the user should use message aggregation,  \rev{or other existing MPI-3.1 semantics}, in order to send as few messages as possible. \rev{With large partition size, however, a higher number of partitions will lead to greater performance benefit, as latency and thread contention become negligible.}
Partitioned communication then delivers the expected benefits of pipelined communication, similar to other existing MPI methods.%

From our work, we estimate that the strength and the weakness of the partitioned communication semantics are in the ease of use of its interface.
The latter leads to suboptimal performance when used with small messages, because of thread contention.
The reason is not new to MPI: to provide a well-performing implementation for both many partitions per thread and many threads, the implementation needs to be able to exploit a user-provided thread context identifier.
Doing so would guarantee no conflict when accessing the resources and the optimal performance in every scenario.
A user worried about performance should therefore use other existing strategies such as \code{MPI\_Comm\_dup} or the more lightweight \code{MPIX\_Stream}, to isolate communications issued from different threads.

Extending partitioned communication to GPU accelerators is an active topic, which is at the center of our future work.
We also plan to further improve our implementation and to remove the need of synchronization between the sender and the receiver during the first iteration.

\begin{acks}
This research was supported by the Exascale Computing Project (17-SC-20-SC), a collaborative effort of the U.S. Department of Energy Office of Science and the National Nuclear Security Administration.
Computational resources have been provided by EuroHPC for the access to MeluXina (\textit{EHPC-BEN-2022B01}).
\end{acks}

\bibliographystyle{ACM-Reference-Format}
\IfFileExists{/Users/tgillis/Dropbox/vortexbib/AllPublications.bib}{%
\bibliography{/Users/tgillis/Dropbox/vortexbib/AllPublications.bib}
}{
\bibliography{pcomm.bib}
}

\appendix
\section{Delay rate}
\label{sec_delay_rate}

\subsection{Definition}
The delay in the pipelined communication pattern comes from the computation time assumed to be proportional to the partition size ($\spart$) and to the average computation rate $\mu$, and the computation noise to follow a normal distribution, $\mathcal{N}\left( 0 \;;\; \sigma  \mu\right)$, whose standard deviation is proportional to $\mu$.

The average computation rate, $\mu$, depends on several factors:
\begin{itemize}
\item the CPU, represented by its frequency, $F$, and the number of \flop{}s per cycles; and
\item the algorithm used, described by the arithmetic intensity (\ai, given in \flop{}\per{}\B), and the communication intensity (\ci), that is, the number of bytes actually sent/received compared with the memory used by the algorithm.
\end{itemize}
We then obtain that
\be
\mu =  \dfrac{\text{\ai{}}}{\text{\ci}} \; \dfrac{1}{\left(8 \; F \right)}
\eed
On the other hand, the noise accumulated during the computations, $\sigma$, depends on two other factors: the algorithmic imbalance in the computations (different branches lead to different computations), $\delta$; and the noise in the system execution\cite{Dosanjh:2021}, $\epsilon$.
Hence we get that $\sigma = (\epsilon + \delta)/2$.
Finally, we obtain that the computation time of a given partition is given by
\be
T_{cmpt} = \mu \; \mathcal{N}\left( 1 \;;\; \dfrac{\epsilon + \delta}{2} \right)
\eed

The delay time in the pipelined communication pattern is the time elapsed between the first partition to be ready and the last one.
Assuming the Gaussian model described earlier, the first partition will be ready after $ \mu \spart (1 - \sigma)$.
The last partition  will be ready once the $\theta$ partitions on the thread have been processed with some delay, $ \mu \spart  \left( \theta + \sqrt{\theta} \sigma \right)$.
The delay time is then obtained as the difference between the two:
\be
D = \gamma_{\theta} \; \spart = \mu \left( \theta + \dfrac{\epsilon + \delta}{2}  (\sqrt{\theta}+1)  -1 \right) \spart
\eec
leading to the definition of the delay rate as being
\be
\gamma_{\theta} = \mu \left( \theta + \dfrac{\epsilon + \delta}{2}  (\sqrt{\theta}+1)  -1 \right)
\eed

\subsection{Numerical examples}
\label{sec_delay_rate_numeric}
\subsubsection{Fourier transform}
A distributed FFT has an $\ai \approx 5$, $\ci = 1$, and $\delta = 0$ (no algorithmic delay)~\cite{Ibeid:2020}.
Assuming a reasonable level of noise ($\epsilon = 0.04$) and $8$ threads, we get a delay rate of $\gamma_1 = 7.1428$ for $\theta=1$, $\gamma_2 = 187.1936$ for $\theta = 2$ and $\gamma_8 = 1263.67$ for $\theta =8$.
The associated gains would then be $\eta = 1.0228$, $\eta = 1.4134$, and $\eta = 1.9748$.

\subsubsection{Finite difference stencil}
For a distributed 3D finite difference stencil, considering one cubic block of data per rank of size $64^3$ and two ghost points, the \ci is $\left(66/64\right)^3 -1 \approx 0.1$. The \ai values of a finite difference stencils are usually around $\approx 1/13$ (4th order) and the $\delta$ can be  high in some applications, $\delta = 0.5$ indicating that some algorithmic branches can lead to $50\%$ more computations.
With $N=8$, we obtain $\gamma_1 = 15.3398$, $\gamma_2 =46.92385411 $, and $\gamma_8 =228.21310932$. The associated gains are then given by $\eta = 1.1060$ for $\theta=1$, $\eta = 1.1718$ for $\theta=1$, and $\eta = 1.2169$ for $\theta=8$.

\end{document}